\documentclass[12pt]{article}
\begin{document}
\begin{center}
25 years ago: the official farewell to the meter
\end{center}
\vskip 0.5cm
On october 21st 1983 took place in S\`evres on the western outskirts of Paris the official funeral of the meter. With it the notion of distance as a physical observable was buried. The office was celebrated in a strict intimacy, but it is difficult to overestimate the loss to physics caused by the disappearance of the meter. Announced in 1915 by Albert Einstein, experimentalists took several decades to agree on the death and to authorise the funeral: ``The 17th Conf\'erence G\'en\'erale des Poids et Mesures decides: the metre is the length of the path travelled by light in vacuum during a time interval of 1/299 792 458 of a second.'' (http://www.bipm.org/en/CGPM/db/17/1/)

This does not exactly read like a death certificate. Still it is one, because time is known to vary from observer to observer and cautious people talk about proper time only. For concreteness suppose we want to measure the distance between the earth and the moon. To make things simple, let us assume that both earth and moon are at rest. On earth we would, since october 1983, measure the time-of-flight for the return trip of a light signal sent to the moon and reflected back to earth. By definition the distance would then be this time of flight divided by two, multiplied by the speed of light. However an astronaut on the moon, repeating the same measurement, would find a different distance.

No distance also means no privileged, ``inertial'', coordinate system and with the meter every theory based on such a coordinate system dies, like Newton's mechanics, Maxwell's electrodynamics and quantum mechanics. And which theory does survive? Historically the first theory to work without the use of inertial coordinates and still today by far the simplest such theory is general relativity. Fortunately it reproduces Newton's mechanics in the limit of low velocities and low gravitational fields. Maxwell's electrodynamics can easily be adapted to general relativity and thereby resuscitated. No such adaption is known today for quantum mechanics. Indeed one of its pillars, Heisenberg's uncertainty relation, refers to position i.e. distance. 

Einstein died in 1955 and could not attend the funeral of the meter. This is sad, not for him, he knew that he was right about the distance being a void concept. It is sad for this generation of physicists because the 1983 General Conference on Weights and Measures passed as a no-event.
\hfil\medskip

\noindent
Thomas Sch\"ucker\\
thomas.schucker@gmail.com

\end{document}